# Threshold for FELWI operation

D.N. Klochkov [1], K.S. Badikyan [2*]

[1] Prokhorov General Physics Institute, Moscow, Russia

[2] National University of Architecture and Construction, Yerevan, Armenia

[*] badikyan.kar@gmail.com

**Abstract:** It is shown that the possibility to create the free electron laser without inversion (FELWI) has a threshold behavior on the field of intensity of amplified wave. In the collective approach, the description of threshold conditions is given. It is shown that the threshold of observation of amplification without inversion is sufficiently high, which hampers essentially the possibility of experimental realization of the FELWI.

## 1. INTRODUCTION

The idea to create the free electron lasers without an inversion (FELWI) was first proposed in [1], and then developed and improved in [2–5]. Specific realizations of the FELWI were proposed and considered in [6, 7]. One of the main points of the FELWI realization schemes is a proposal to use the non-collinear propagations of the electron beam and the amplified radiation. In the conventional free electron lasers (FEL) and the strophotrons these schemes are well known and have been discussed for a long time [8-48]. As applied to the FELWI with two undulators, the main idea is that at the non-collinear interaction of laser and electron beams after first undulator, there is a scattering of electrons over transversal velocities and hence of angles, and this scattering is directly related with the increase in the electron energy. Therefore, the selection of electrons over directions in the undulator gap is equivalent to the selection over the energies.

This mechanism can operate only if the angular spread $\alpha$ resulting from the interaction of electrons with the field of undulator and amplified wave is greater than the natural velocity dispersion over directions in the electron beam $\Delta\alpha_{beam}$. Virtually, $\Delta\alpha_{beam}$ cannot be less than $10^{-4}$ rad. The condition $\alpha > \Delta\alpha_{beam}$ leads to the origination of the realization threshold of the FELWI either by the intensity of laser radiation or by the electron beam density.

In the present work, the threshold of origination of amplification effect without the inversion in the proposed FEL schemes to achieve this goal is evaluated. The analysis is performed within the framework of the multiparticle description covering both the Compton and the Raman amplification modes in the FEL.

## 2. The noncollinear geometry

The theoretical constructs of [arxiv] suggest the infinite electron and light beams. In reality, both are limited in the transverse direction. For non-collinear electron and laser beams the latter circumstance results to the finite region of their interaction. The length at which the light amplification takes place in the medium of the beam of electrons is (see the figure)

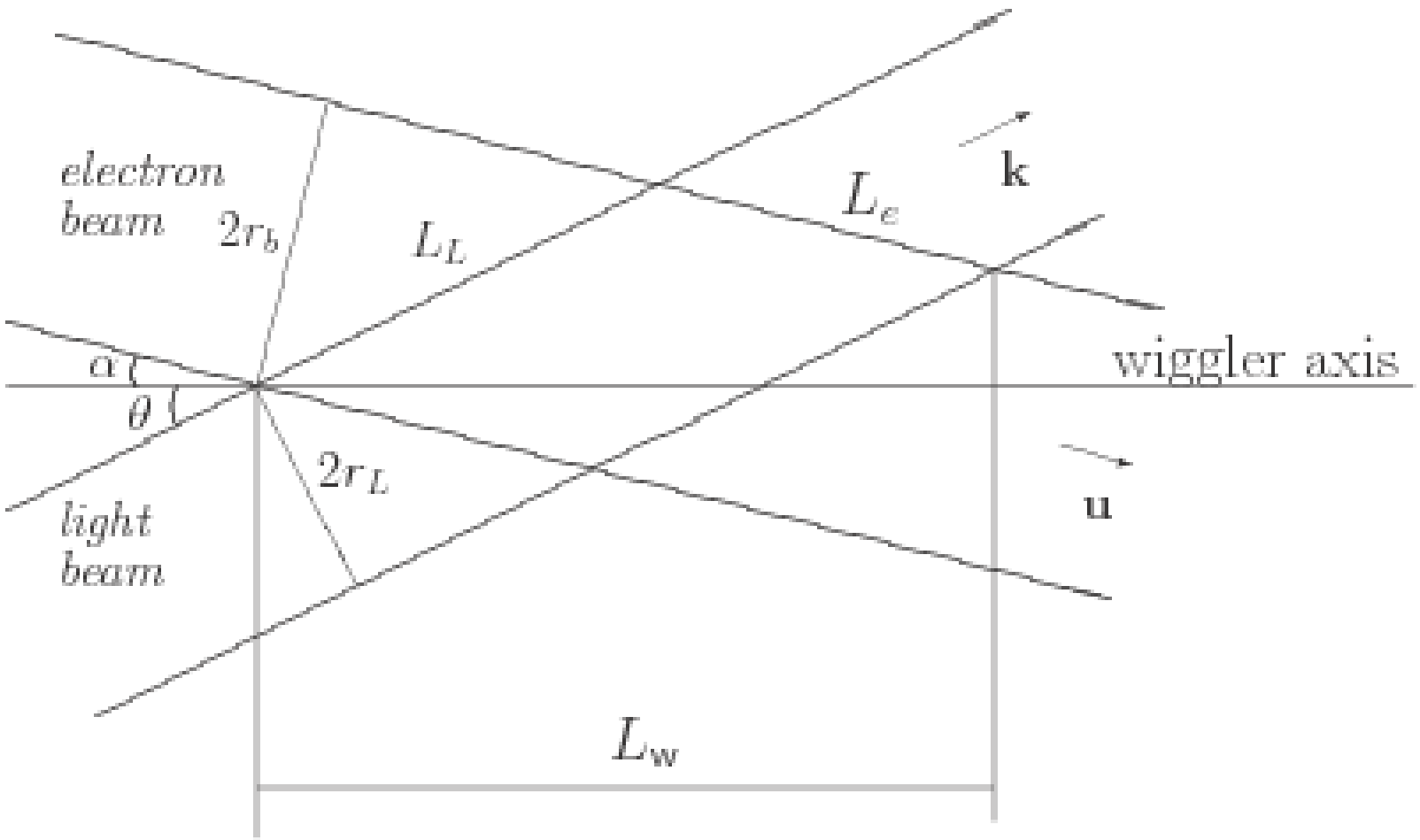

Fig.1. The scheme in $xz$ -plane of undulator with non-collinear arrangement.

$$L_L = \frac{2r_L}{\sin(\alpha+\theta)}. \tag{14}$$

Here $2r_b$ is the width of the electron beam in the $xz$-plane. In turn, the length over which the laser field does the work over the electron is

$$L_e = \frac{2r_L}{\sin(\alpha+\theta)}. \tag{15}$$

where $2r_L$ is the width of the electron beam in the $xz$-plane. The operating length of wiggler is defined by

$$L_w = L_e \cos\alpha + L_L \cos\theta \approx \frac{2(r_L + r_b)}{\sin(\alpha+\theta)}. \tag{16}$$

It makes no sense to increase the length of the wiggler more than $L_w$. In order to increase $L_w$, the quantity $\alpha+\theta$ should be decreased and the width of the electron beam $2r_b$ should be increased. The latter should be done also in virtue of the following estimates.

The gain of the laser field by the wave amplitude is equal to

$$\frac{A_{\text{out}}}{A_{\text{in}}} = \exp\left(k'' L_L\right). \tag{17}$$

Acording to [arxiv], $k'' \sim \omega_b^{\nu}$. For the Thomson (single-particle) regime of instability $\nu = 2/3$ and $\nu = 4/3$, for the Raman (collective) regime of instability $\nu = 1/2$ and $\nu = 3/2$. For the cylindrical form of the beam of electrons, when the current is constant, one has the estimate $\omega_b \sim r_b^{-1}$ and therefore $k'' \sim \omega_b^{\nu} \sim r_b^{-\nu}$. Since $L_L \gg r_b$ one obtains the estimate for the gain

$$\frac{A_{\text{out}}}{A_{\text{in}}} = \exp\left(const \times r_b^{1-\nu}\right). \tag{18}$$

When $\nu < 1$, we have the monotonically increasing function on the width $r_b$ of the beam of electron and the monotonically decreasing in the opposite case $\nu > 1$. Thus, at $\nu < 1$ a wide beam should be used and at $\nu > 1$ (super high-current electron beams) one should use the narrow beam.

## 3. Threshold for Amplification Without Inversion (AWI)

Consider here the estimations as applied to the FELWI. As mentioned above, the FELWI underlying mechanism can work only if the angular spread α resulting from the interaction of electrons with the field is greater than the natural spread $\Delta\alpha_{\text{beam}}$ of the electron beam by directions. This fact leads to the origination of the threshold for the power of laser radiation.

Regarding to its unperturbed motion $\mathbf{r} = \mathbf{r}_0 + \mathbf{u}t$ in the *xz*-plane, the electrons of beam undergo the oscillations in this plane, with the variations of velocity equal to in the linear approximation [44]

0 ω

$$\delta\mathbf{v}_{\perp} = K^2 \frac{c^2}{\gamma_0^3} \frac{\beta_1 \mathbf{k} - \frac{\omega}{c^2}\beta_2 \mathbf{u}}{D_b} a e^{i\xi_0 - i\Delta_\omega t} + \text{c.c.}. \tag{19}$$

Here $a = a_+ / A_0$ is the dimensionless amplitude of the laser field, $\Delta_\omega = \omega - (\mathbf{ku})$, $\xi_0 = \mathbf{k}_0 \mathbf{r}_{\perp 0}$, $\mathbf{r}_{\perp 0}$ is the initial coordinate in the plane *xz*. Coefficients $\beta_1$ and $\beta_2$ are equal:

$$\beta_1 = \gamma_0\left(\omega - (\mathbf{k}_0\mathbf{u})\right) - \frac{\omega_b^2 (\mathbf{k}_0\mathbf{u})}{(k_0 c)^2},$$
$$\beta_2 = \gamma_0\left(\omega - (\mathbf{k}_0\mathbf{u})\right) - \frac{\omega_b^2}{\omega}. \tag{20}$$

The equation of the trajectory (19) is obtained for a monoenergetic beam which has an infinite size,

and thereby it interacts with the field endlessly. However, using this equation, one can obtain the necessary estimations, which will be carried for the most interesting case from the point of view of experiment – for the case of the single-particle regime of the gain. Assuming that an electron enters the laser field at the time instant $t = 0$ and, during the flight time $t = L_e / u$ it deflects from its original direction by an angle $\Delta\alpha$ which depends on the phase of entry into the field according to the $\cos\xi_0$ law, we obtain the maximum value

$$\Delta\alpha_{\max} = K^2 \frac{c^2}{\gamma_0^2} \frac{k_0}{u^2} \sin(\alpha + \theta) \frac{e^{ik''L_e} - 1}{k''}. \tag{21}$$

In the case of a weak amplification on the length $k''L_e << 1$, the rotation angle does not depend on the angle $\alpha + \theta$ , nor from the gain coefficient *k* '' (and hence on the beam current):

$$\Delta\alpha_{\max} = \frac{2K^2 k_0 r_L}{\gamma_0^2} a. \tag{22}$$

As expected, the value of $\Delta\alpha_{\max}$ coincides with the estimate given in the works [34, 41]. The expression (22) should be considered as the lower limit of the maximum possible deviation. The excess $\Delta\alpha_{\max}$ of natural spread $\Delta\alpha_{\text{beam}}$ gives the threshold value for the amplitude of the laser field and for its intensity. Rewrite (22) in terms of the total laser power $W = \frac{c}{4}(k_0 r_L)^2 \left|a_+\right|^2$ , namely by an expression defining the power threshold:

$$W > \frac{c}{8}\left(\frac{mc^2}{e}\right)^2 \frac{\left(\Delta\alpha_{\text{beam}}\right)^2 \gamma_0^4}{2K^2}. \tag{23}$$

This gives the numerical value

$$W > 1.1\times10^9 \frac{\left(\Delta\alpha_{\text{beam}}\right)^2 \gamma_0^4}{2K^2} (\text{watt}). \tag{24}$$

For the following values of the parameters [7] $\gamma_0 = 15$, K = 0.635 , and $\Delta\alpha_{\text{beam}} = 5\times10^{-4}$ rad and we obtain the threshold value $W > 1.8\times10^7\,\text{W}$ . The resulting threshold power exceeds the power of the 164 laser field at which the saturation occurs. To use the laser in a linear amplification regime $\Delta\alpha_{\text{beam}}$ should be reduced. When operating at the very border of the saturation region $\sim 10^5 - 10^6\ \text{W}/\text{см}^2$, the formula (23) gives an estimate $\Delta\alpha_{\text{beam}} \sim 10^{-6}$ rad.

This estimate coincides with the results of [34] obtained by an another (single-particle) approach. Note that for the stable operation of the FELWI the value $\Delta\alpha_{\max}$ should exceed the natural spread of the electron beam by directions $\Delta\alpha_{beam}$ by an order of magnitude. It is doubtful that the natural angular divergence of an electron beam can be achieved in the accelerator significantly less than $10^{-6}$ rad.

## 4. CONCLUSION

The influence of the velocity spread of the electron beam by directions on the work of the FELWI is investigated. Within the multiparticle approach, the threshold value of maximal angular spread by velocities of beam electrons is obtained. It is shown that the threshold for the angular spread corresponds to the presence of a threshold by intensity (or power) of the laser field. The obtained threshold value the power of the laser radiation (23) is an upper limit under condition of a small gain. As follows from the above estimates, the realization of the FELWI in the single-particle regime with the small gain $k''L << 1$ encounters the big problems that result in the practical impossibility to its realization in this regime, in the version which was considered above. There are two ways out of this situation. Firstly, the use of the Raman amplification regime. As numerical simulations have shown [7], the FELWI idea has great possibilities in this regime. Secondly, the use of strong amplification at the length $k''L >> 1$ in the Thompson regime. In both cases there is a need for the use of the high density beams.